\documentclass[preprint,12pt,numbers,numbers,sort&compress]{elsarticle}

\usepackage{amssymb}
\usepackage{amsmath}
\usepackage{xcolor}  
\usepackage{CJKutf8}

\journal{Fusion Engineering Design}

\begin{document}

\begin{frontmatter}




\title{Effect{s} of {external} magnetic field ripple on FRC equilibrium} 


\author[label1]{Zitong Qu}
\author{Ping Zhu\corref{cor1}\fnref{label1,label2}}
\ead{zhup@hust.edu.cn}
\author{Zhipeng Chen\corref{cor1} \fnref{label1}}
\ead{zpchen@hust.edu.cn}
\author[label3]{Haolong Li} 
\cortext[cor1]{Corresponding author.}

\affiliation[label1]{organization={State Key Laboratory of Advanced Electromagnetic Technology, International Joint Research Laboratory of
Magnetic Confinement Fusion and Plasma Physics, School of Electrical and Electronic Engineering},
            addressline={Huazhong University of Science and Technology}, 
            city={Wuhan},
            postcode={430074}, 
            country={China}}
            
\affiliation[label2]{organization={Department of Nuclear Engineering and Engineering Physics, University of Wisconsin-Madison},
            addressline={Madison}, 
            city={Wisconsin},
            postcode={53706}, 
            country={United States of America}}
            
\affiliation[label3]{organization={Department of Physics, School of Science},
            addressline={Tianjin University of Science and Technology},
            city={Tianjin},
            postcode={300457},
            country={China}}
\begin{abstract}
The two-dimensional equilibrium of Field-Reversed Configuration (FRC) plasma {in presence} of an external ripple magnetic field is {computed to show} the emergence of multiple magnetic axes for hollow equilibr{ium current profiles}. {A}n increase in ripple amplitude reduces the hollow{ness} threshold required for the development of multiple magnetic axes. {For an intermediate range of the ripple axial period}, the formation of multiple magnetic ax{e}s becomes the most likely. {T}he ripple's {radial} {extension} and the curvature of the axial field {are the} critical factors underlying the non-monotonic effect of ripple's {axial} period. {When} the {axial} period increases {from the lower range}, the ripple’s radial {extension} gradually {grows and enhances} the chance of forming multiple magnetic axes. {O}nce the ripple’s radial {extension covers} the entire radial {domain}, {further increasing the ripple axial period} decreas{es} the ripple field {curvature, which} becomes the dominant factor {for} lowering the hollowness threshold {for the formation of multiple magnetic axes}.


\end{abstract}




\begin{keyword}


Field reversed configuration \sep equilibrium \sep multi{ple} magnetic axes \sep hollow {current} profile
\end{keyword}

\end{frontmatter}

\section{Introduction}
\label{sec1}
Field Reversed Configurations (FRCs) are high-beta compact toroidal magnetic confinement fusion devices {with minimal} toroidal magnetic field \cite{tuszewski1988nf,steinhauer2011pop,woodruff2008jofe}. {Equilibrium construction} of FRC is crucial {to its} transport and stability {analysis}. The pressure profiles identified in previous studies can be classified into three types : peaked \cite{morse1970pof,sparks1980pf,clemente1984pof,wang2024pop,galeotti2011pop}, flat \citep{berk1981pof,spencer1982pof,suzuki1985jpsj}, and hollow \cite{cobb1993pofpp,steinhauer2009pop,steinhauer2014pop,lee2020nf,ma2021nf} {, and Steinhauer defined the profile index $h$ \cite{steinhauer1992pof} to quantify the degree of profile hollowness}. Experiments indicate that  {FRC} equilibri{a} predominantly adopt hollow profiles \cite{steinhauer1992pof,steinhauer2009pop,tuszewski2011fst}, and an increased degree of hollowness enhance{s} {the} FRC stability \citep{cobb1993pofpp,steinhauer1994pop, kanno1995jpsj, slough1995pop}. {As a result}, {studies} have focused on {the} hollow profile {to} better address the requirements of experimental analysis and device design. 


Non-uniform coil spacing in experiments results in increased magnetic field ripple, which impedes the formation and maintenance of high performance FRC \citep{peng2022nf,zhang2023fed}. {This study aims} {to} investigate the impact{s} of magnetic field ripple on {the} hollow equilibria{,} with particular {attentions to the} ripple amplitude and {its axial} period. {Equilibrium computations} demonstrate that external coil ripple {can} lead to the formation of multi-axis magnetic structures in hollow equilibria. Moreover, both ripple amplitude and 
spatial period distinctly influence the maximum hollowness {required for} the emergence of multiple magnetic axes. These findings may help {design of} mitigati{on} schemes for {the avoidance} of multi-ax{e}s in {FRC} experiment{s}.


This paper is organized as follows: The method{s} for computing {the} ripple fields and the equilibrium are described in Section 2. Section 3 presents two-dimensional equilibrium solutions that {in presence of} ripple {fields} and examines the impact of ripple {axial} period on the development of multi-axis. Finally, Section 4 concludes the study {along with a discussion}.


\section{{Methodology}}
\label{sec2}

\subsection{Ripple field calculation}
\label{subsec21}

The external magnetic field is {composed of} a magnetic mirror field and a ripple field. {These fields are generated by external colis, which are} represented {as} current filament{s}, with each coil group consist{ing} of a single turn to {signify} the ripple effect. The flux function at the vacuum vessel boundary $\psi_w$ is expressed as:


\begin{equation}
\psi_w=\sum_{i=1}^{N_c}G(R_w,Z_w,{R^{c}_{i},Z^c_i)I^c_{i}}+\sum_{j=1}^{ {N_m}}G(R_w,Z_w,{R^m_{j},Z^m_{j})I^m_{j}} \label{psiw}
\end{equation}
where \(G\) represents the Green's function, \(R_{w}\) and \(Z_{w}\) denote the coordinates {of} the vacuum vessel wall, {\( R^{m}_{j} \)} and  {\( Z^{m}_{j} \)} the mirror coils, {\(R^{c}_{i}\) and \(Z^{c}_{i}\)} the ripple field coils, {and} {\(I^{m}_{j}\)} {and {\(I^{c}_{i}\)}} the mirror {and ripple} coil current{s respectively}. The index \(i\) identifies each individual coil, while \(N_{c}\) denotes the total number of ripple field coil sets. Throughout the analysis, the magnetic mirror field {configuration is kept unchanged}, including the number of coils {(\( N_{m} \))}, their spatial {locations (\( R^{m}_{j} \) and \( Z^{m}_{j} \))}, and the corresponding currents {(\( I^{m}_{j} \))}. 



The magnetic field components are  {obtained using}
 {$B_R = -{R}^{-1} {\partial \psi}/{\partial Z}$} and  {$B_Z = {R}^{-1} {\partial \psi}/{\partial R}$}. The magnetic field ripple amplitude is defined as {$B_{\text{ripp}}(R) = {\max(B_{Z,\max} - B_{Z,\min})}/{B_{Z,\text{ave}}}$},
where \( B_{Z,\max} \), \( B_{Z,\min} \){, and \( B_{Z,\text{ave}} \)} denote the local maximum, the adjacent minimum, and the average of {$B_Z$ along the axial direction at each radius $R$}, respectively. The ripple axial period \( L_{p} \) is defined as the average distance between consecutive local maxima. The flux ripple amplitude {\( \psi_{\text{ripp}} \)} is defined similarly. To examine the effects of varying ripple amplitude and axial period, the average axial magnetic field at the vacuum vessel is held constant {when the ripple amplitude or the axial period is each individually varied alone in turn}{, b}y adjusting the coil current {\(I^{c}_{i}\)},{the} spacing ${L_d}$, and {its}  {radial coordinate} {$R^c_i$}.


\subsection{Equilibrium calculation}
\label{subsec22}
The macroscopic {parameters for the} characteristics of FRC plasma equilibria {include the} elongation (\(E\)), {and} the average{d} $\beta=p/(B_Z^2/2\mu_0)$ on the midplane $(Z=0)$ inside the separatrix, {among others}. {In addition,} Steinhauer \citep{steinhauer1992pof} introduced the current profile index \(h\), a single parameter that characterizes the equilibrium in terms of its toroidal current profile {as follows}


\begin{equation}
h\equiv {\frac{(j_{\phi}/R)_{O}}{\langle j_{\phi}/R\rangle}}
\end{equation}
{where ${(j_{\phi}/R)_{O}}$ is the {"current" at} the O-point (magnetic axis), and} $\langle j_{\phi}/R \rangle$ denotes the mean value within the midplane separatrix \citep{steinhauer2009pop}


\begin{align}
\langle j_\phi/R\rangle\equiv(1/\pi R_s^2)\int_0^{R_w}(j_\phi/R)2\pi RdR 
\end{align}
When employing the Grad–Shafranov (GS) equation, $h$ also represents the ratio of the first derivative of pressure at point $O$ to its average value. When $h < 1$, the {current} exhibits a hollow {profile}; conversely, for $h > 1$, it assumes a peaked profile. The specific case of $h = 1$ corresponds to a flat {current} profile, as exemplified by Hill's vortex equilibrium \citep{steinhauer2011pop,spencer1982pof}. Figure \ref{fig3} illustrates the current density and pressure profiles for various values of $h$.


{A} {model pressure profile is} selected {for} its relatively simple form, {and its} flexibility in adjusting the hollowness {of} current density profile \citep{takahashi2004pop}  {as well as} its ability to account for {the} plasma characteristics {outside} separatrix.


\begin{align}
\left.p(\psi)=\left\{\begin{matrix}{p_{open}+b_{0}+b_{0}b_{1}\psi+\frac{1}{2}b_{0}b_{1}^{2}\psi^{2}+b_{2}\psi^{3},\psi\leq0}\\
{p_{open}+b_{0}exp(b_{1}\psi),\psi>0}\end{matrix}\right.\right.\label{pmodel}
\end{align}
where $b_0$, $b_1$, and $b_2$ are constants, and $p_{\text{open}}$ represents the minimum pressure outside the separatrix. The magnetic flux $\psi$ is {set to be negative} inside the separatrix. {T}he pressure and its first and second derivatives {are continuous} at the separatrix, which can prevent the introduction of surface currents and the finite Larmor radius effect \citep{steinhauer2014pop}, thereby facilitating subsequent stability analysis. The pressure profile inside the separatrix is modeled as a cubic polynomial of magnetic flux, a simple form that allows for variation in the constants $b_1$ and $b_2$ {to model various} hollow current density profiles. Outside the separatrix, the pressure decays exponentially with magnetic flux $\psi$, consistent with {those} observed experimentally.


Equilibrium is obtained by numerically solving the GS equation using the finite difference method, with boundary conditions specified by $\psi(R_w(Z),Z)=\psi_w$ {and} Eq{.}\thinspace\eqref{psiw}. Neumann condition is applied at both ends, i.e., ${\partial\psi}/{\partial Z}=0${. A} uniform grid is employed \citep{ma2021nfGSEQ}. The solver framework follows the GSEQ code \citep{ma2021nfGSEQ}. To avoid bifurcated solutions and accelerate convergence, a global constraint is imposed on the plasma current at the midplane, $I_{1D}=2\pi\int_0^{R_w}J_\phi RdR$. The pressure is expressed as $p=C{p}(\psi)$, where $C=wC_\mathrm{old}+(1-w)C_\mathrm{old}  ({I_{1D}}/{I_\mathrm{mid}})$, and $I_\mathrm{mid}$ is the plasma current per unit length at the midplane. Unlike GSEQ, this current is specified as a fixed value rather than being calculated from a one 
dimensional equilibrium. To adjust the hollowness of the two-dimensional equilibrium profile, the constant $b_{2}$ is {varied, which} only affects the pressure profile inside the separatrix, as shown in Eq.\thinspace\eqref{pmodel}.


\section{Result and discussion}
\label{sec3}

\subsection{Equilibrium {in presence of} ripple}
The hollow {current} equilibrium exhibits multiple magnetic axes (Figure \ref{fig5}(a)) when the ripple magnetic field is applied, whereas the peaked {current} equilibrium maintains a single magnetic axis (Figure \ref{fig5}(b)). The presence of multiple magnetic axes is identified {with} the multiple extrema of the magnetic flux function along the {magnetic} axis \( {R} = {R}_o \), {which} correspond{s} to the minimum {location of the} flux  {function}. The hollow equilibrium help{s} stabilize the FRC instability, {and} the stabilizing effect becom{es} more pronounced as the {hollowness} increases (i.e., as $h$ decreases)  \citep{cobb1993pofpp,kanno1995jpsj,steinhauer1994pop}. However, the {presence} of multiple magnetic axes {may} introduce {additional} impact on instability. Therefore, it is {important} to {find out} the specific range of equilibrium solutions with multiple magnetic axes in hollow equilibria {in presence} of magnetic field ripple.


To facilitate comparison, \(h'=1-h\) is defined to quantify the {hollowness of} the current density profile. A positive \( h' \) (\( h' > 0 \)) indicates a hollow pressure profile, whereas a negative \( h' \) (\( h' < 0 \)) corresponds to a peaked pressure profile, with \( h' = 0 \) representing a flat pressure profile. The critical hollow{ness} for the occurrence of a multiple magnetic axis equilibrium is defined as \( h'_t \), {above which} the magnetic field ripple {can introduce multiple magnetic axes to the equilibrium}. 


\subsection{Results}

The first step is to investigate the effect of the boundary axial magnetic field ripple amplitude on the hollowness threshold for the occurrence of multiple magnetic axes in hollow equilibria. The average axial magnetic field at the vacuum wall is held constant at \( B_{Z, \text{ave}} = 0.04 \, \text{T} \), {and} the axial period of the axial magnetic field ripple, {\( {L_p} = 0.04 \, \text{m} \)}, {is also} fixed. The threshold for the hollowness, \( h'_t \), {above} which multiple magnetic axes {appear}, is computed which decreases {at} a slower rate as the ripple amplitude increases  {(Figure \ref{fig6})}. This suggests that a larger ripple amplitude requires a lower hollowness for the occurrence of multiple magnetic axes, thereby {tends to cause} the formation of a multi-axis equilibrium.


The following examines the influence of {the period of the boundary magnetic field ripple} on the threshold for the formation of multiple magnetic axes in hollow equilibria. The average axial magnetic field at the vacuum wall is maintained at \( B_{Z, \text{ave}} = 0.04 \, \text{T} \), {and the ripple period is varied {under a series of ripple amplitude of }}0.4, 0.6, 0.8, and 1.0 (Figure \ref{fig7}). {T}he {threshold} \( h'_t \)  first decreases and then increases {with respect to the ripple period}, {and} the {ripple period for} the minimum {$h'_t$} varies {with} the ripple amplitude. {This suggests the ripple field is {the} most effective in bringing multiple magnetic ax{e}s to equilibrium only within an intermediate range of ripple period.}


\subsection{Discussion}
 To {understand} the non-monotonic {influence} of the {ripple} period, two factors are considered: the radial extension and the curvature of the ripple field. {Figure \ref{fig8} shows the radial distribution of the ripple amplitudes for various ripple periods. The decay of the ripple amplitude  toward the cylindrical axis indicates that the effect of the ripple field to the plasma equilibrium is not global and limited near the source, such as the external coils. The distributions for various ripple periods in Figure \ref{fig8} are different, which means the range of influence of the ripple field varies with the ripple period. The ratio of the ripple field between the axial and wall is calculated to represent the extension of the ripple field and shown under different ripple period in the Figure \ref{fig9}. As the ripple period increases, this ratio also increases, indicating a stronger ripple effect in the core region.}


As the axial period of the ripple increases, the radial extension of its effects expands, making the occurrence of multiple magnetic axes more likely. During this process, the hollowness threshold decreases as the range of the magnetic field ripple effects expands {as the fall of the $h'_t$ with short $L_p$ in the figure \ref{fig7}.}  Once the ripple radial effects cover the entire radial computational domain, further increases in ripple 
extension {would} have a {limited} impact on the hollowness threshold.




{A}s the ripple period {increases}, the overall curvature of the ripple field decreases correspondingly. The axial magnetic field at the boundary can be approximated by the following cosine function: \(B_Z = \left( 1 - (B_{\text{ripp}}(R_w)/2 ) \right) \,\allowbreak\) \linebreak \( \cos \left( (2 \pi/{{L_p}}) Z \right) \,\allowbreak B_{Z, \text{ave}}\). {The curvature is}
{
\begin{align}
{
\vec{\kappa} \;=\;
\frac{2k^{2}R \left( \sin^{2}(kZ) + 2 \right) \cos(kZ)}{D^{3}}
\left(
\frac{2 \cos(kZ)}{D} \,\overrightarrow{\mathbf{e}_{R}}
-
\frac{kR \sin(kZ)}{D} \,\overrightarrow{\mathbf{e}_{Z}}
\right)
}
\end{align}
}
{\!where \(k = 2\pi / L_p\), \(D = \sqrt{k^2 R^2 \sin^2(kZ) + 4 \cos^2(kZ)}\). The average magnetic field curvature {\(\left\langle \, |\vec{\kappa}| \, \right\rangle\) } along the boundary (\(R=R_w, Z \in [-Z_w, \, Z_w]\)) is evaluated and found to} decrease as the axial period of the magnetic field ripple increases (Figure \ref{fig10}). This suggests a reduction in the coil ripple effect, which is further indicated by the subsequent increase in the hollowness threshold value after the minimum point.


\section{Summary}
\label{sec4}
The two-dimensional equilibrium of FRC plasma {in presence of} a ripple external magnetic field is calculated and found to exhibit multiple magnetic axes when the equilibrium {current profile} is hollow. The effects of the magnetic field's ripple amplitude and axial period on the emergence of multiple magnetic axes are subsequently investigated. The results show that as the ripple amplitude increases, the maximum {hollowness} required for the occurrence of multiple magnetic axes in the hollow equilibrium decreases, thereby increasing the likelihood of a multi-axis equilibrium. The effect of the ripple's axial period on {the hollowness threshold for} the development of multiple magnetic axes is non-monotonic: when the {ripple} period is either {small} or large, the multiple magnetic axes {are less likely to form}.


An analysis of both the radial extension and the curvature of the {ripple} field provides insight into the non-monotonic effect of the {ripple} period. As the axial period of the ripple increases, the radial extension of the ripple field gradually expands, {facilitating} the emergence of multiple magnetic axes. Once the influence range covers the entire radial computational region, further increases in the ripple's radial extension have a diminish{ing} effect on reducing the hollowness threshold. Additionally, as the curvature of the axial field decreases with {the further} increasing {of} ripple period, {the multiple axes }{become more difficult to form and the hollowness threshold goes up again 
after the ripple spans the entire radial domain}.


Therefore, for hollow profile commonly observed in experiments, which exhibit stabilizing effects, it is essential to reduce coil ripple to maintain a single magnetic axis equilibrium {with hollow current profile}. Future work will investigate the impact of multiple magnetic axis on {the} stability {of the} hollow equilibria.


\section{Acknowledgments}
The authors are very grateful for the help of the HFRC team in the State Key Laboratory of Advanced Electromagnetic Engineering and Technology of China. This work was supported by  the National MCF Energy R\&D Program of China (Grant No.2019YFE03050004),  the National Key Research and Development Program of China (Grant No.2017YFE0301804 and No.2017YFE0301805), the Hubei International Science and Technology Cooperation Project under Grant No. 2022EHB003, and the U.S. Department of Energy (Grant No.DEFG02-86ER53218). The computing work in this paper was supported by the Public Service Platform of High Performance Computing by Network and Computing Center of HUST.


\begin{graphicalabstract}
\end{graphicalabstract}

\begin{figure}[htbp]

\centering
\includegraphics[width=\textwidth]{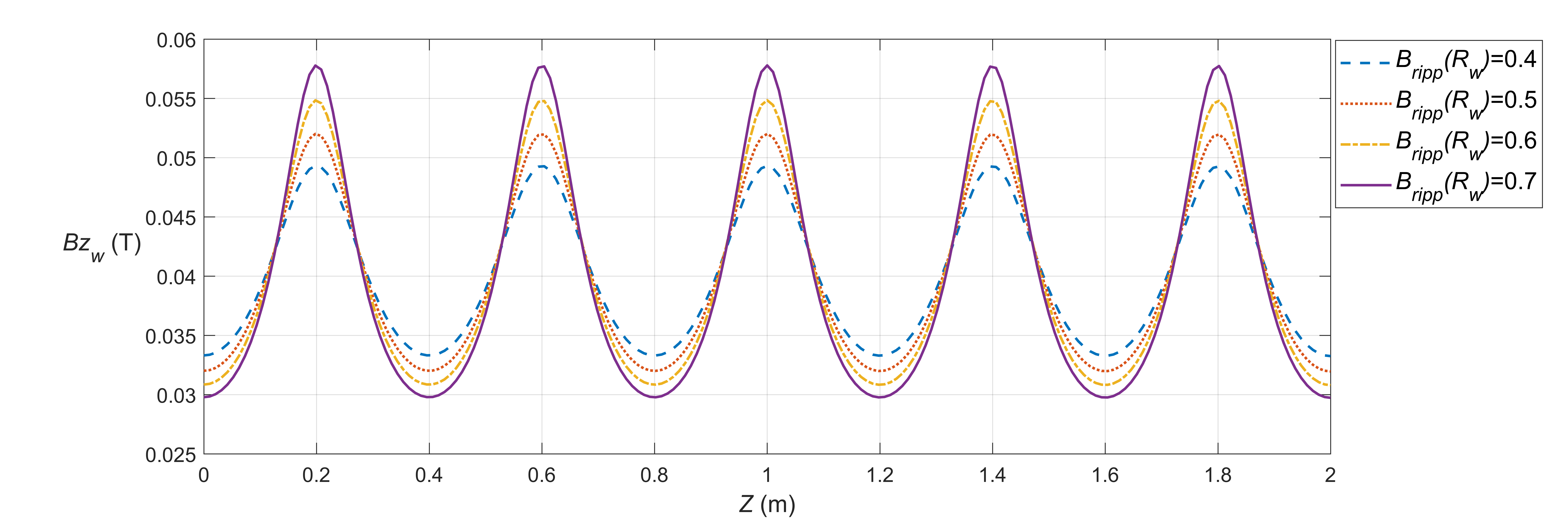}
\caption{Axial magnetic field{s $B_{Z_w}$ at wall as functions of the axial coordinate $Z$} with the same axial period and {various} ripple amplitudes. Since the axial magnetic field and magnetic flux at the wall are both symmetrical about the midplane, only half of the {axis is} shown.
}\label{fig1}
\end{figure}

\begin{figure}[htbp]
\centering
\includegraphics[width=\textwidth]{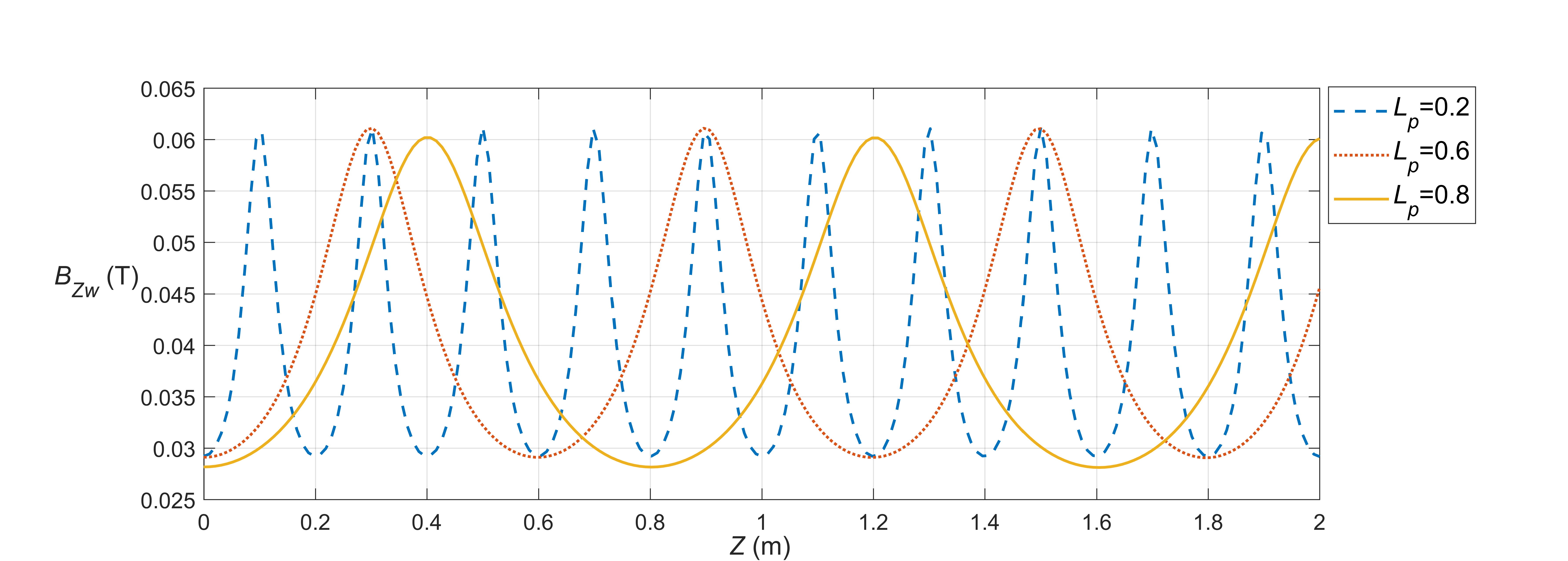}
\caption{Axial magnetic field{s $B_{ Z_w}$ at wall as functions of the axial coordinate $Z$} with the same amplitudes and {various} ripple period{s}. Since the axial magnetic field and magnetic flux at the wall are both symmetrical about the midplane, only half of the {axis is} shown.
}\label{fig2}
\end{figure}

\begin{figure}[htbp]
\centering
\includegraphics[width=\textwidth]{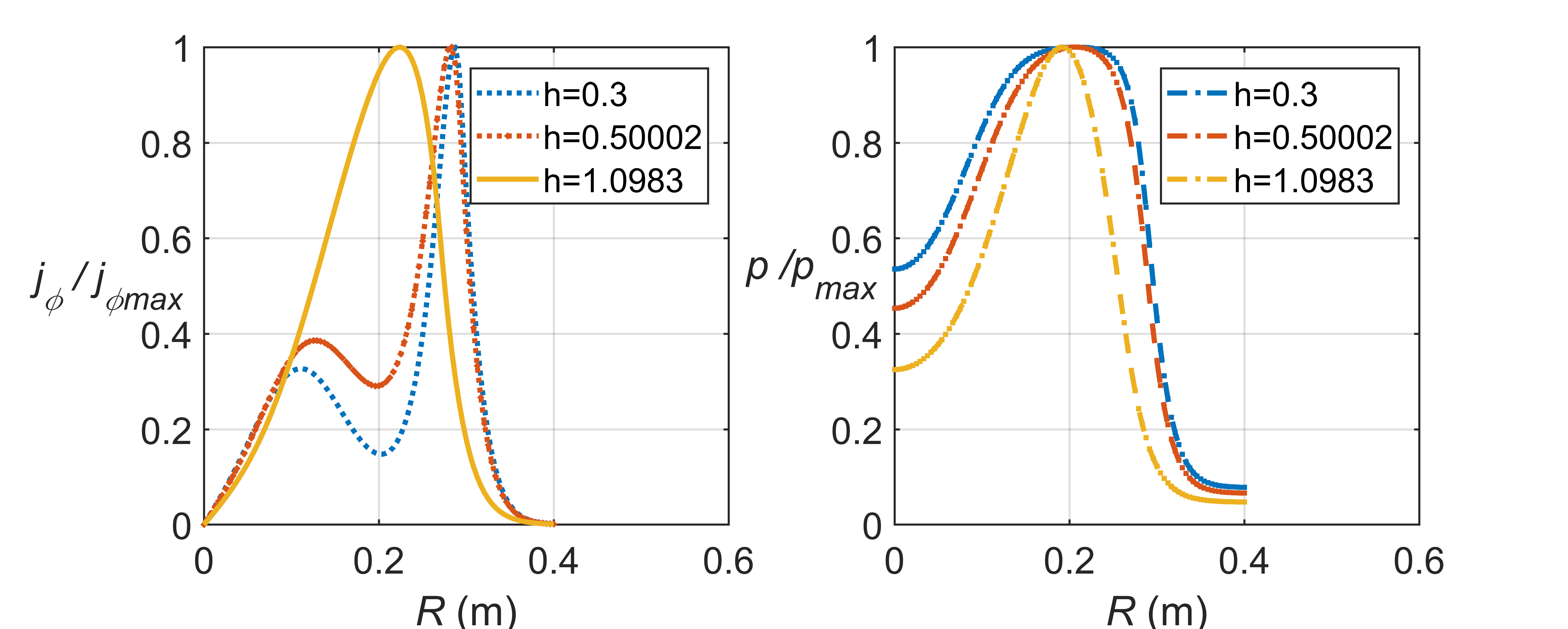}
\caption{The current density and pressure profile {along radius $R$} for {various} current profile ind{ice}{s}.
}\label{fig3}
\end{figure}

\begin{figure}[htbp]
\centering
\includegraphics[width=\textwidth]{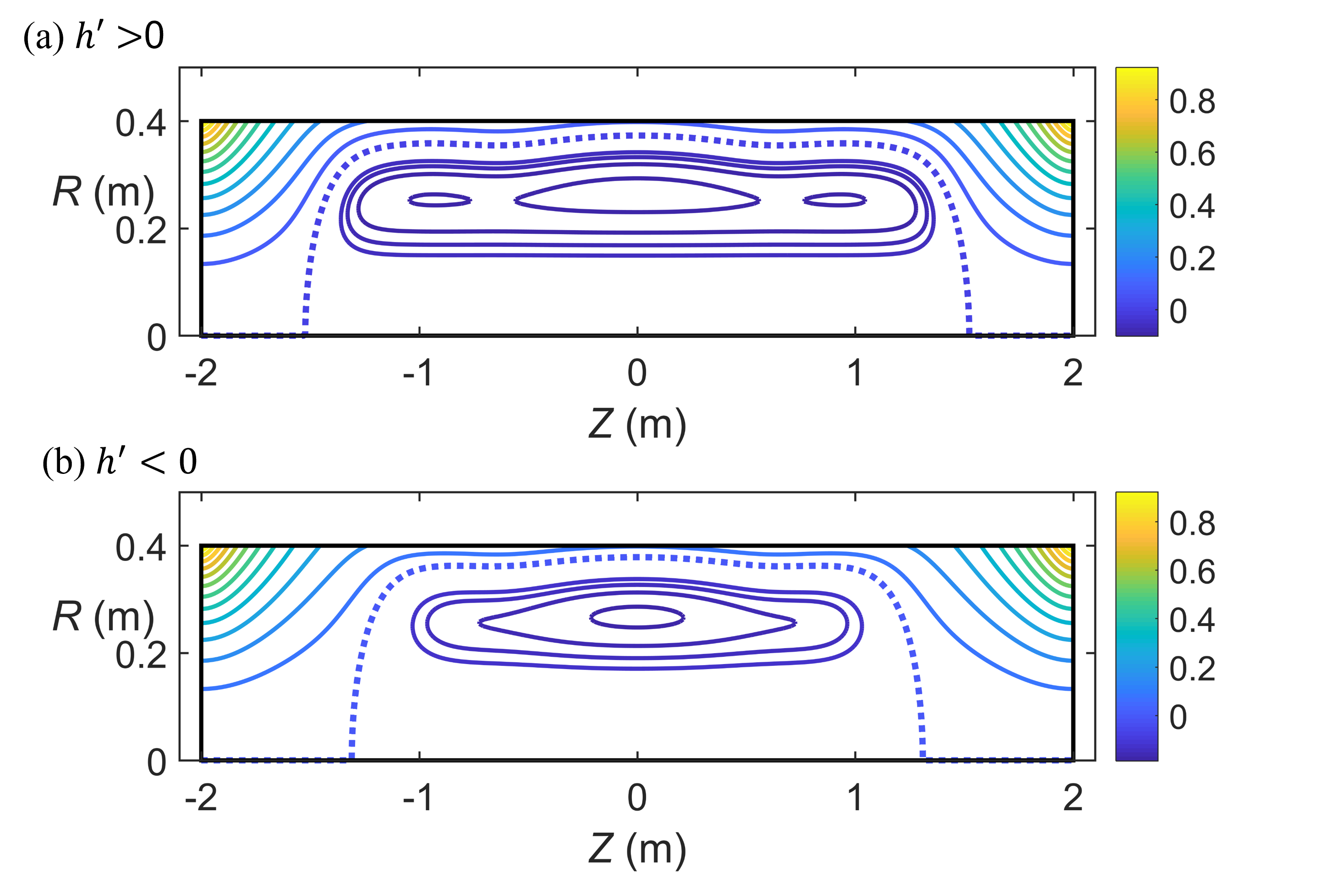}
\caption{{Flux functions of (a) m}ulti-axis hollow equilibrium and {(b) }single{-}axis peaked equilibrium calculated {for} the same coil boundary {condition.}
}\label{fig5}
\end{figure}

\begin{figure}[htbp]
\centering
\includegraphics[width=\textwidth]{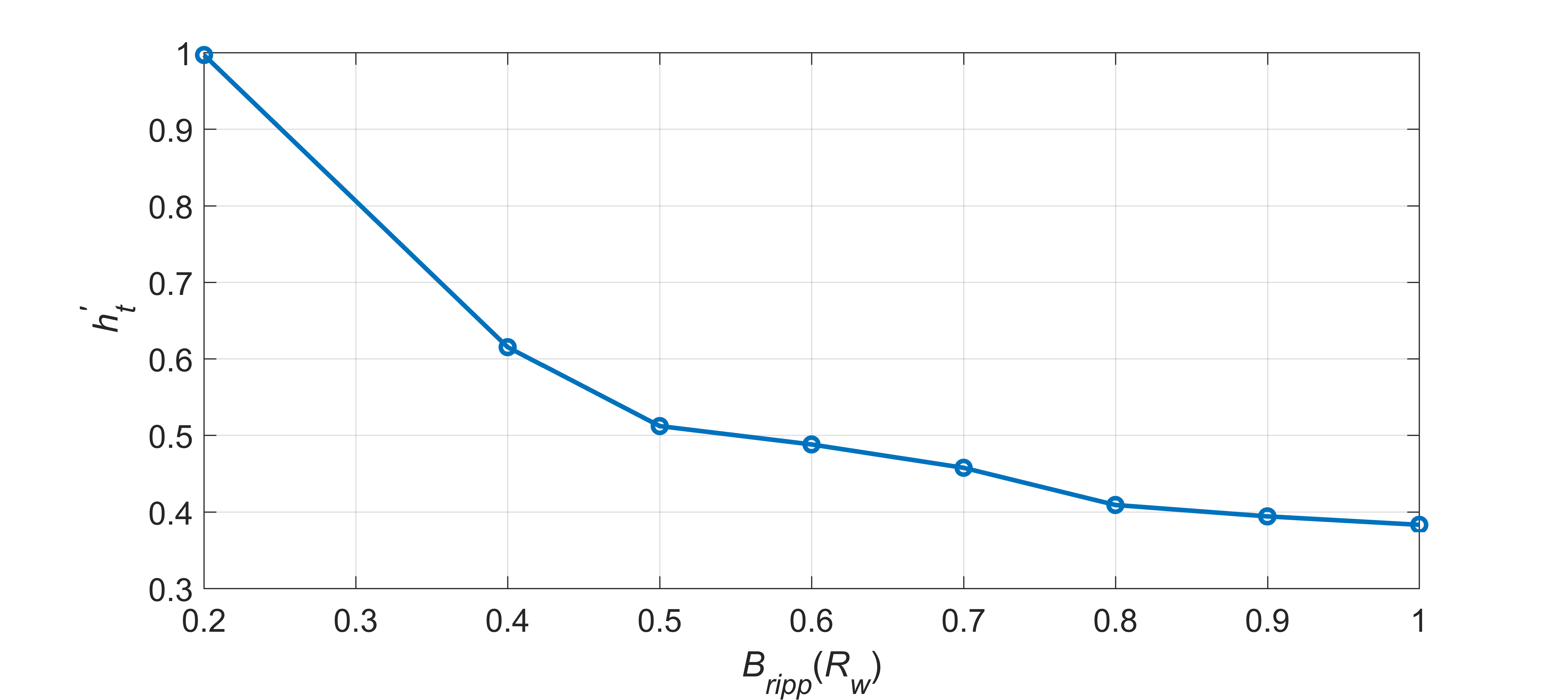}
\caption{The variation of \( h'_{\text{t}} \) with respect to the amplitude of magnetic field ripple.
}\label{fig6}
\end{figure}

\begin{figure}[htbp]
\centering
\includegraphics[width=\textwidth]{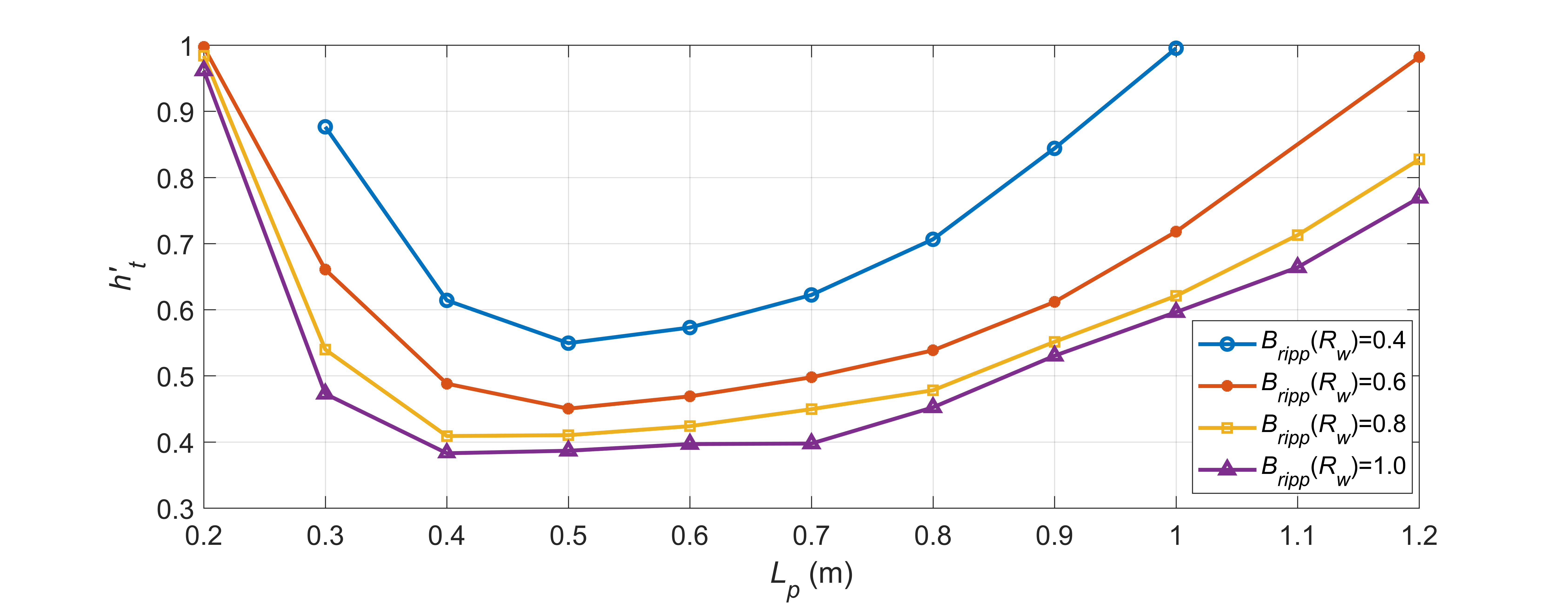}
\caption{The variation of \( h'_{\text{t}} \) with respect to the {axial} period of magnetic field ripple.
}\label{fig7}
\end{figure}

\begin{figure}[htbp]
\centering
\includegraphics[width=\textwidth]{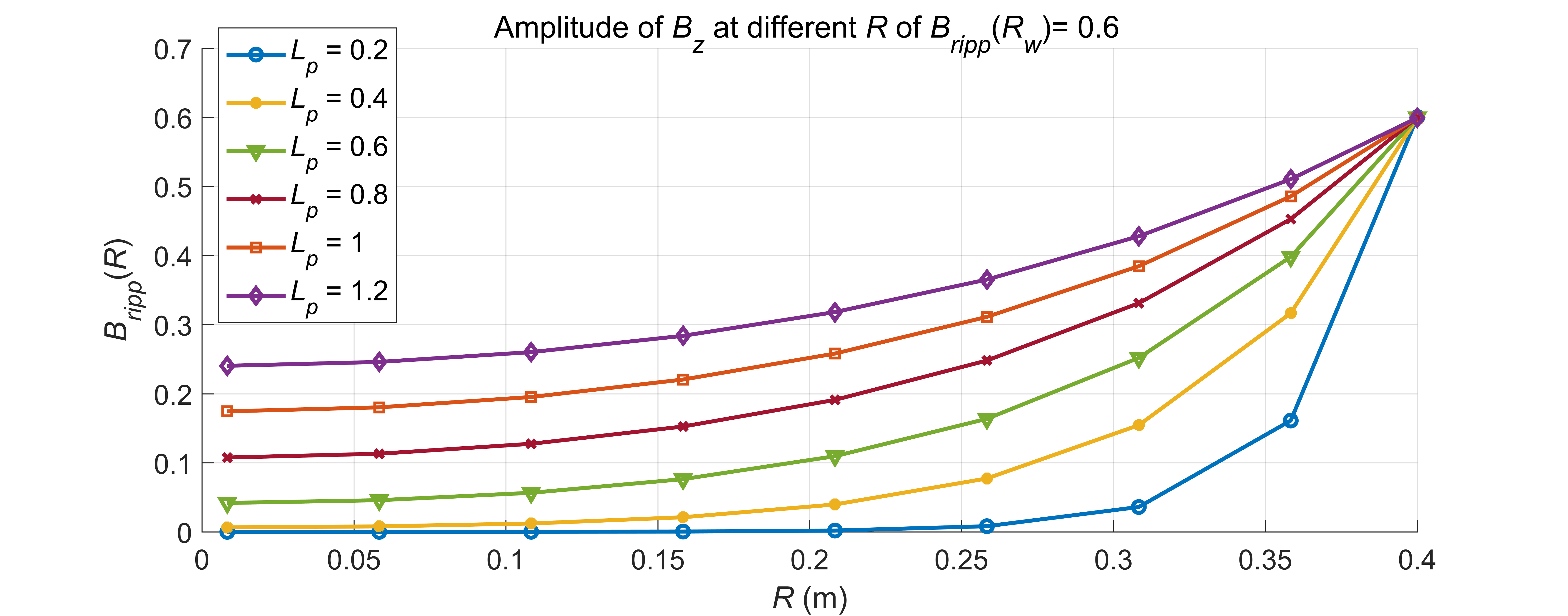}
\caption{Amplitudes of ripple fields {as functions of radius for various ripple periods.}
}\label{fig8}
\end{figure}

\begin{figure}[htbp]
\centering
\includegraphics[width=\textwidth]{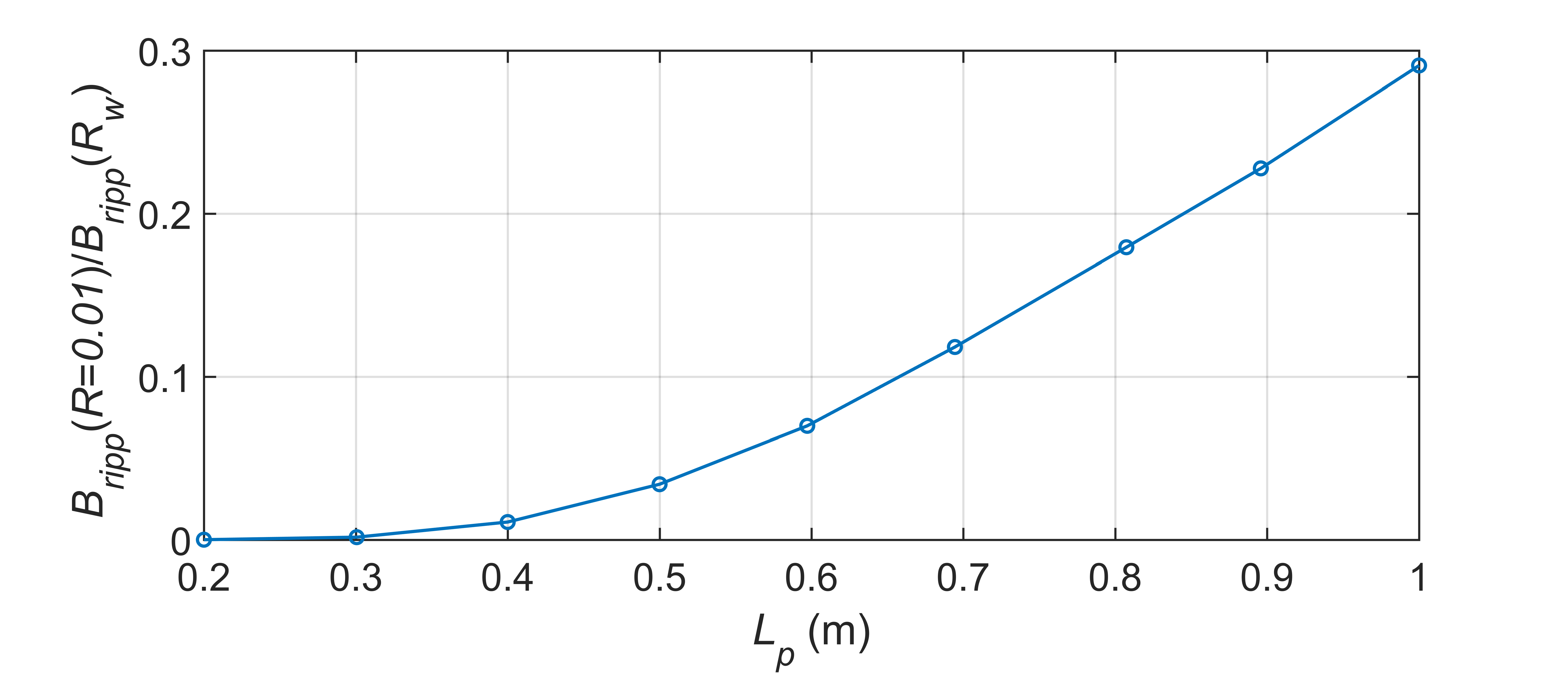}
\caption{{The ratio of ripple amplitude near the axis to the wall under various periods.}
}\label{fig9}
\end{figure}

\begin{figure}[htbp]
\centering
\includegraphics[width=\textwidth]{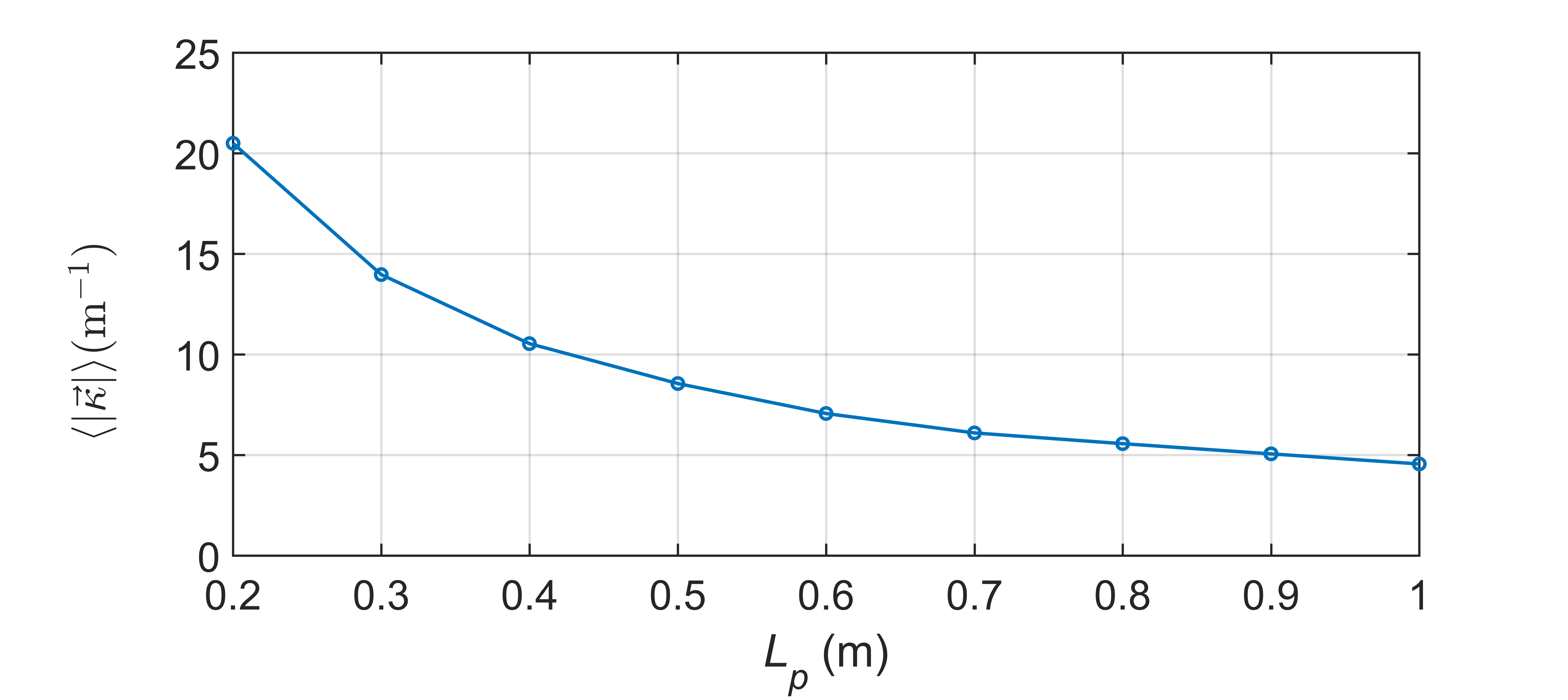}
\caption{Curvature of the boundary axial magnetic field at the midplane {as a function of} the axial period of the boundary magnetic field ripple.
}\label{fig10}

\end{figure}

\clearpage 

\bibliographystyle{elsarticle-num-names} 
\bibliography{ref}

@article{tuszewski1988nf,
  title={Field reversed configurations},
  author={Tuszewski, Michel},
  journal={Nuclear Fusion},
  volume={28},
  number={11},
  pages={008},
  year={1988},
  publisher={IOP Publishing}
}

@article{steinhauer2011pop,
  title={Review of field-reversed configurations},
  author={Steinhauer, Loren C},
  journal={Physics of Plasmas},
  volume={18},
  number={7},
  year={2011},
  pages={070501},
  publisher={AIP Publishing}
}

@article{woodruff2008jofe,
  title={Technical survey of simply connected compact tori ({CTs}): Spheromaks, {FRC}s and compression schemes},
  author={Woodruff, S},
  journal={Journal of fusion energy},
  volume={27},
  pages={134--148},
  year={2008},
  publisher={Springer}
}

@article{steinhauer1992pof,
  title={Profile consistency in equilibria of field-reversed configurations},
  author={Steinhauer, Loren C and Ishida, Akio},
  journal={Physics of Fluids B},
  volume={4},
  number={3},
  pages={645--650},
  year={1992}
}

@article{morse1970pof,
  title = {Rigid {{Drift Model}} of {{High-Temperature Plasma Containment}}},
  author = {Morse, R. L.},
  year = {1970},
  journal = {Physics of Fluids},
  shortjournal = {Phys. Fluids},
  volume = {13},
  number = {2},
  pages = {531},

  urldate = {2020-12-21},
  langid = {english},
}

@article{clemente1984pof,
  title = {A Class of Rotating Compact Tori Equilibria},
  author = {Clemente, R. A. and Farengo, R.},
  year = {1984},
  journal = {Physics of Fluids},
  shortjournal = {Phys. Fluids},
  volume = {27},
  number = {4},
  pages = {776},
  urldate = {2021-04-12},
  langid = {english}
}

@article{wang2024pop,
  title = {A Gyrokinetic Simulation Model for {{2D}} Equilibrium Potential in the Scrape-off Layer of a Field-Reversed Configuration},
  author = {Wang, W. H. and Wei, X. S. and Lin, Z. and Lau, C. and Dettrick, S. and Tajima, T.},
  year = {2024},
  journal = {Physics of Plasmas},
  shortjournal = {Physics of Plasmas},
  volume = {31},
  number = {7},
  pages = {072507},
  urldate = {2024-07-28}
}

@article{cobb1993pofpp,
  title={Profile stabilization of tilt mode in a field-reversed configuration},
  author={Cobb, John W and Tajima, T and Barnes, Daniel C},
  journal={Physics of Fluids B: Plasma Physics},
  volume={5},
  number={9},
  pages={3227--3238},
  year={1993},
  publisher={American Institute of Physics}
}

@article{steinhauer1994pop,
  title={Ideal stability of a toroidal confinement system without a toroidal magnetic field},
  author={Steinhauer, Loren C and Ishida, A and Kanno, R},
  volume={1},
  journal = {Physics of Plasmas},
  number={5},
  pages={1523--1528},
  year={1994},
  publisher={American Institute of Physics}
}

@article{kanno1995jpsj,
  title={Ideal-magnetohydrodynamic-stable tilting in field-reversed configurations},
  author={Kanno, Ryutaro and Ishida, Akio and Steinhauer, Loren C},
  journal={Journal of the Physical Society of Japan},
  volume={64},
  number={2},
  pages={463--478},
  year={1995},
  publisher={The Physical Society of Japan}
}

@article{steinhauer2014pop,
  title = {Two-Dimensional Interpreter for Field-Reversed Configurations},
  author = {Steinhauer, Loren},
  year = {2014},
  date = {2014-08-01},
  journal = {Physics of Plasmas},
  volume = {21},
  number = {8},
  pages = {082516},
  urldate = {2023-10-22}
}

@article{ma2021nf,
  title = {Two-Parameter Modified Rigid Rotor Radial Equilibrium Model for Field-Reversed Configurations},
  author = {Ma, H.J. and Xie, H.S. and Bai, Y.K. and Cheng, S.K. and Deng, B.H. and Tuszewski, M. and Li, Y. and Zhao, H.Y. and Chen, B. and Liu, J.Y.},
  year = {2021},
  journal = {Nuclear Fusion},
  shortjournal = {Nucl. Fusion},
  volume = {61},
  number = {3},
  pages = {036046},
  urldate = {2021-03-15}
}

@article{ma2021nfGSEQ,
  title={A new tool {GSEQ-FRC} for two-dimensional field-reversed configuration equilibrium},
  author={Ma, HJ and Xie, HS and Deng, BH and Bai, YK and Cheng, SK and Li, Yang and Chen, Bin and Tuszewski, Michel and Zhao, HY and Liu, JY},
  journal={Nuclear Fusion},
  volume={61},
  number={8},
  pages={086006},
  year={2021},
  publisher={IOP Publishing}
}

@article{lee2020nf,
  title = {Generalized Radial Profile of Field-Reversed Configurations Based on Symmetrical Properties},
  author = {Lee, K.Y.},
  year = {2020},
  date = {2020-04-01},
  journal = {Nuclear Fusion},
  shortjournal = {Nucl. Fusion},
  volume = {60},
  number = {4},
  pages = {046010},
  urldate = {2021-01-01}
}

@article{galeotti2011pop,
  title = {Plasma Equilibria with Multiple Ion Species: {{Equations}} and Algorithm},
  author = {Galeotti, L. and Barnes, D. C. and Ceccherini, F. and Pegoraro, F.},
  journal = {Physics of Plasmas},
  year = {2011},
  volume = {18},
  number = {8},
  pages = {082509},

}

@article{slough1995pop,
  title={Transport, energy balance, and stability of a large field-reversed configuration},
  author={Slough, JT and Hoffman, AL and Milroy, RD and Maqueda, R and Steinhauer, LC},
  journal={Physics of Plasmas},
  volume={2},
  number={6},
  pages={2286--2291},
  year={1995},
  publisher={AIP Publishing}
}

@article{steinhauer2009pop,
  title={Equilibrium paradigm for field-reversed configurations and application to experiments},
  author={Steinhauer, Loren C and Intrator, TP},
  journal={Physics of Plasmas},
  volume={16},
  number={7},
  year={2009},
  pages={072501},
  publisher={AIP Publishing}
}

@article{peng2022nf,
  title={Simulation on formation process of field-reversed configuration},
  author={Peng, Yue and Yang, Yong and Jia, Yuesong and Rao, Bo and Zhang, Ming and Wang, Zhijiang and Wang, Hongyu and Pan, Yuan},
  journal={Nuclear Fusion},
  volume={62},
  number={6},
  pages={066037},
  year={2022},
  publisher={IOP Publishing}
}

@article{sparks1980pf,
  title={Interchange stability of axisymmetric field reversed equilibria},
  author={Sparks, L and Finn, JM and Sudan, RN},
  journal={Phys. Fluids},
  volume={23},
  number={3},
  year={1980},
  publisher={Laboratory of Plasma Studies, Cornell University, Ithaca, New York 14853}
}

@article{berk1981pof,
  title = {Analytic Field‐reversed Equilibria},
  author = {Berk, Herbert L. and Hammer, James H. and Weitzner, Harold},
  year = {1981},
  journal = {The Physics of Fluids},
  shortjournal = {The Physics of Fluids},
  volume = {24},
  number = {9},
  pages = {1758--1759},
  urldate = {2024-12-02}
}

@article{spencer1982pof,
  title = {Free Boundary Field‐reversed Configuration ({{FRC}}) Equilibria in a Conducting Cylinder},
  author = {Spencer, Ross L. and Hewett, Dennis W.},
  year = {1982},
  journal = {The Physics of Fluids},
  shortjournal = {The Physics of Fluids},
  volume = {25},
  number = {8},
  pages = {1365--1369},
  urldate = {2024-12-04},
}

@article{suzuki1985jpsj,
  title = {Two {{Dimensional Field Reversed Equilibria}} with {{Plasma}} Outside the {{Separatrix}}},
  author = {Suzuki, Kiyomitsu},
  year = {1985},
  journal = {Journal of the Physical Society of Japan},
  shortjournal = {J. Phys. Soc. Jpn.},
  volume = {54},
  number = {6},
  pages = {2155--2159},
  publisher = {The Physical Society of Japan},
}

@article{zhang2023fed,
  title = {Analysis and Design of In-Vessel Magnetic Compression Coil System for {{HFRC}}},
  author = {Zhang, Qinglong and Rao, Bo and Yang, Yong and Zhang, Ming and Lv, Yiliang and Peng, Tao and Wang, Zhijiang and Pan, Yuan},
  journal = {Fusion Engineering and Design},
  year = {2023},
  volume = {194},
  pages = {113751},
}

@article{takahashi2004pop,
  title={Losses of neutral beam injected fast ions due to adiabaticity breaking processes in a field-reversed configuration},
  author={Takahashi, Toshiki and Inoue, Koji and Iwasawa, Naotaka and Ishizuka, Takashi and Kondoh, Yoshiomi},
  journal={Physics of Plasmas},
  volume={11},
  number={6},
  pages={3131--3140},
  year={2004},
  publisher={American Institute of Physics}
}

@article{tuszewski2011fst,
  title={Combined {FRC} and mirror plasma studies in the {C-2} device},
  author={Tuszewski, Michel and Smirnov, Artem and Deng, BH and Dettrick, SA and Song, Y and Andow, R and Barnes, D and Binderbauer, MW and Bui, DQ and Clary, R and others},
  journal={Fusion Science and Technology},
  volume={59},
  number={1T},
  pages={23--26},
  year={2011},
  publisher={Taylor \& Francis}
}

\begin{highlights}
\item Research highlight 1
\item Research highlight 2
\end{highlights}






\end{document}